\documentclass[11pt]{JHEP3}

\usepackage{amsmath,latexsym,amssymb}
\usepackage{psfrag}
\usepackage[latin1]{inputenc}
\usepackage{subfigure}
\usepackage{nicefrac}
\usepackage{bbm}
\usepackage[stable]{footmisc}
\usepackage{fancyhdr}
\usepackage[pdftex]{graphicx}
\usepackage{psfrag}
\usepackage{fancybox}
\usepackage{ifthen}
\usepackage{epsfig}
\usepackage[errorshow]{tracefnt}
\usepackage{multirow}
\newcommand{\be}{\begin{equation}}
\newcommand{\ee}{\end{equation}}
\newcommand{\ben}{\begin{displaymath}}
\newcommand{\een}{\end{displaymath}}
\newcommand{\bea}{\begin{eqnarray}}
\newcommand{\eea}{\end{eqnarray}}

\newcommand{\bean}{\begin{eqnarray*}}
\newcommand{\eean}{\end{eqnarray*}}
\newcommand{\beqs}{\begin{eqnarray}}
\newcommand{\eeqs}{\end{eqnarray}}
 
\renewcommand{\d}{\ensuremath{\textnormal{d}}}

\newcommand\W{\widetilde{W}}

\newcommand\bbone{\ensuremath{\mathbbm{1}}}

\newcommand\beal{\begin{align}}

\newcommand{\eq}[1]{\begin{equation}#1\end{equation}}

\newcommand{\spl}[1]{\begin{split}#1\end{split}}

\newcommand\benu{\begin{enumerate}}
\newcommand\eenu{\end{enumerate}}
\newcommand\bit{\begin{itemize}}
\newcommand\eit{\end{itemize}}

\newcommand\cm{\mathcal{M}}

\newcommand{\boxedeq}[1]{
\begin{equation}
\fbox{
\rule[0.7cm]{0pt}{0pt}
$#1$
\rule[-0.45cm]{0pt}{0pt}
}
\end{equation}
}

\title{Rigid supersymmetric theories in 4d Riemannian space}

\author{Henning Samtleben${}^{\diamondsuit}$ and Dimitrios Tsimpis${}^{\clubsuit}$
  \\

  \begin{itemize}

\item  Universit\'e de Lyon, Laboratoire de Physique, UMR 5672, CNRS et ENS de Lyon,\\
46 all\'ee d'Italie, F-69364 Lyon CEDEX 07, France \\
Institut Universitaire de France
  
\item Universit\'{e} de Lyon\\
UMR 5822, CNRS/IN2P3, Institut de Physique Nucl\'{e}aire de Lyon\\ 
4 rue Enrico Fermi,  
F-69622 Villeurbanne Cedex, France
  \end{itemize}

\bigskip
 E-mail:
\email{henning.samtleben@ens-lyon.fr}, \email{tsimpis@ipnl.in2p3.fr} }

\abstract{We consider rigid supersymmetric theories in four-dimensional Riemannian spin manifolds. 
We build the Lagrangian directly in Euclidean signature from the outset, 
keeping track of potential boundary terms. 
We reformulate the conditions for supersymmetry as 
a set of conditions on the torsion classes of a suitable $SU(2)$ or trivial $G$-structure. 
We illustrate the formalism with a number of examples 
including supersymmetric backgrounds with non-vanishing Weyl tensor. 
}

\preprint{$\,$}
\begin{document}

\newpage

\section{Introduction}\label{s1}

Several exact results have by now been obtained for supersymmetric gauge theories, such as  the computation 
of indices, partition functions and Wilson loops, providing in many cases checks of highly non-trivial dualities. Such calculations rely for the most part on localization techniques -- which in their turn rely on 
the theory being rigid supersymmetric in some curved, in general, background. In four dimensions, recent studies include supersymmetric theories on $S^4$ \cite{a}, $S^1\times S^3$ \cite{b,c,d}, $S^1\times S^3/\mathbb{Z}_k$ \cite{e,f}, $S^1\times L(p,q)$ \cite{g} and $AdS_4$ \cite{Adams:2011vw,kuzenko}. In view of the success of this program, it would be interesting to extend this list of four-dimensional spaces to more general backgrounds.

In the present paper we will focus on rigid supersymmetric theories in four-dimensional Riemannian spin manifolds. In other words the four-dimensional  background in which the theory lives is assumed to be equipped with a positive-definite 
metric of Euclidean signature. 
A systematic approach to the study of rigid supersymmetry in four-dimensional curved space has recently been initiated in \cite{fs}. As follows from the analysis of \cite{fs} the condition for a theory to be rigidly supersymmetric in a given background reduces to the requirement for the existence of a pair of Killing spinors\footnote{The Killing spinor equations are given in (\ref{kse}) below; our use of the term `Killing spinor' is more general than is sometimes assumed in the literature.} on that background. Here, `background' refers to the bosonic fields of the minimal off-shell supergravity multiplet in four dimensions, i.e.\ to the choice of the metric $g_{mn}$ as well as background fields $b_m, M, \bar{M}$ which appear as parameters in the globally supersymmetric action and supersymmetry transformation rules.

From the technical standpoint the analysis of such Killing-spinor equations in four-dimen\-sional Riemannian space $\mathcal{M}_4$ can be performed using a suitable $G$-structure. For generic backgrounds the pair of Killing spinors defines a local trivialization of the structure group of 
$T\mathcal{M}_4$, i.e.\ a trivial $G$-structure. For certain backgrounds one of the spinors is allowed to vanish identically, in which case the other,  non-vanishing, spinor defines a local $SU(2)$ structure.
In either case, the $G$-structure is given explicitly by locally constructing a set of forms in terms of bilinears of the Killing spinors. The Killing spinor equations can then be reexpressed as a set of constraints on the torsion classes of the $G$-structure.

In this paper, we pursue this approach for a systematic study of the Killing-spinor equations in four-dimensional Riemannian space.
More specifically, in the case of a trivial $G$-structure we reformulate the 
conditions for unbroken supersymmetry as the set of constraints on 
the torsion classes given in (\ref{217}) below. Equivalently, 
 we derive a set of necessary and sufficient conditions for unbroken rigid supersymmetry, given in eqn.~(\ref{ns}):   
Given a Riemannian four-manifold $\mathcal{M}_4$ and a trivial $G$-structure   such that eqs.~(\ref{ns}) are satisfied, all background fields are uniquely determined and the theory is (at least) $\mathcal{N}=1$ rigid supersymmetric.
Similarly, in the case of an $SU(2)$-structure we reformulate the 
conditions for unbroken supersymmetry as the set of constraints on 
the torsion classes given in (\ref{su2torsion}). 

We begin our analysis in section \ref{s2} by formulating the rigid supersymmetric theory directly in Euclidean signature. The Lagrangian is given in (\ref{l}) and was constructed from scratch, without reference to any Wick rotation.  Up to boundary terms which we compute explicitly, the Lagrangian is invariant under transformations (\ref{st}), where the  supersymmetry parameters $\zeta$, $\xi$ obey the pair of coupled Killing-spinor equations (\ref{kse}): this is our definition of rigid $\mathcal{N}=1$ supersymmetry. More generally, the background possesses $\mathcal{N}\geq 1$ supersymmetry if and only if the space of solutions to the   linear system of differential equations (\ref{kse}) is $\mathcal{N}$-dimensional.

In section \ref{s22} we work out the reformulation of the supersymmetry conditions in terms of the trivial $G$-structure; in section \ref{ssu2} we do the same in the case of an $SU(2)$ structure. 
We illustrate the formalism using several examples in section \ref{examples}. We start in section \ref{k3} 
with the example of a K3 surface, while sections \ref{ts}, \ref{th} treat background geometries of the form $T^d\times S^{d-4}$ and $T^d\times H^{d-4}$ respectively. In all the cases except for $T^2\times S^2$ the solution extends globally on $\mathcal{M}_4$, i.e.\ all these backgrounds possess global Killing spinors. In the case of $T^2\times S^2$ the background fields develop singularities at 
the poles. Treating the poles as a boundary, taking into consideration the total derivatives, the `bulk' action can be shown to be supersymmetric. The case $T^2\times \mathcal{M}_2$ for an arbitrary 
two-dimensional Riemannian manifold $\mathcal{M}_2$ is treated in section \ref{sec:gen} and the local existence of Killing spinors is shown. Section \ref{sec:genff} presents an example of a conformally flat four-manifold.

The backgrounds $T^2\times S^2$ and $T^2\times H^2$ have a non-vanishing Weyl tensor and do not belong to the list of examples considered explicitly in \cite{fs}. Moreover for the $S^1\times S^3$, the $S^1\times H^3$, as well as the example in section \ref{sec:genff}, all of which have a vanishing Weyl tensor, we will present solutions to the Killing spinor equations 
which violate the conditions in \cite{fs} and only allow for $\mathcal{N}<4$ supersymmetries.

Our Euclidean spinor conventions are further explained in appendix \ref{sec:spinors}.

\section{Rigid supersymmetry in Riemannian space}\label{s2}

We will work in a Riemannian space $\mathcal{M}_4$ parameterized by 
coordinates $x^m$, $m=1,\dots, 4$. The  starting point of the 
supersymmetry analysis is the set of Killing spinor equations of~\cite{fs}. In our spinor conventions, which are further 
explained in appendix \ref{sec:spinors}, these read:
\begin{equation}\begin{split}
\nabla_m\zeta+M\gamma_m\xi+2b_m\zeta+b^n\gamma_{nm}\zeta&=0\\
\nabla_m\xi-\bar{M}\gamma_m\zeta-2b_m\xi-b^n\gamma_{nm}\xi&=0
~,
\end{split}
\label{kse}\end{equation}
in a given background defined by the metric and the fields $(b^m,M,\bar{M})$.
In the following we take $b_m$ to 
be a complex one-form on $\mathcal{M}_4$ and $M$, $\bar{M}$ to be independent complex scalars. Note that in the case of Minkowski signature, taking $b_m$ to be  imaginary and setting $\bar{M}=M^*$ the second line above is the complex conjugate of the first line. However 
in Euclidean signature the spinors $\zeta$, $\xi$ are Weyl pseudoreal of opposite chirality;  contrary to the case in Minkowski signature, they are independent and can never be related to each other by complex conjugation.

The globally supersymmetric Lagrangian can be obtained by evaluating the off-shell supergravity Lagrangian
of~\cite{Cremmer:1982en} on a background that allows for solutions of (\ref{kse}) and setting the
gravitino fields to zero. The passage to Euclidean space can be performed by proper
Wick rotation, see e.g.~\cite{vanNieuwenhuizen:1996tv}; instead for the following we have constructed the 
Lagrangian from scratch. Up to terms quartic in the fermions, the resulting Lagrangian is given by
\begin{eqnarray}\label{l}
{\cal L} &=&
-\left(\tfrac16 R
+4 M\bar{M} -4 b_m b^m \right) K
-g_{i\bar{\jmath}} \left(
\partial_m \phi^i \partial^m \bar{\phi}^{\bar{\jmath}} + F^i \bar{F}^{\bar{\jmath}} \right)
\nonumber\\
&&{}
+ g_{i\bar{\jmath}} \,\tilde{\psi}^{\bar{\jmath}}\, \gamma^m 
(\nabla_m \psi^i +\Gamma^i_{jk} \psi^j \partial_m \phi^k 
)
-\tfrac12 K_{ij\bar{\jmath}}\,\bar{F}^{\bar{\jmath}}\,\tilde{\psi}^i \psi^j 
-\tfrac12 K_{i\bar{\imath}\bar{\jmath}}\,{F}^{i}\,\tilde{\psi}^{\bar{\imath}} \psi^{\bar{\jmath}} 
\nonumber\\
&&{}
+2\left(
M K_i F^i + \bar{M} K_{\bar\imath} \bar{F}^{\bar\imath} \right)
+MK_{ij} \tilde\psi^i \psi^j +
\bar{M} K_{\bar\imath\bar\jmath} \tilde\psi^{\bar\imath}\psi^{\bar\jmath}
\nonumber\\
&&{}
+b^m\left(
2K_i \,\partial_m\phi^i -  2K_{\bar\imath}\, \partial_m\bar{\phi}^{\bar\imath} 
- K_{i\bar{\jmath}}\, \tilde\psi^{\bar\jmath} \gamma_m \psi^i
\right)
\nonumber\\
&&{}
+ 3\left(\bar{M} W +  M \bar{W} \right) -\tfrac12\left( F^i W_i +  \bar{F}^{\bar\imath} \bar{W}_{\bar\imath} \right)
-\tfrac14\left( W_{ij} \,\tilde\psi^i\psi^j +  \bar{W}_{\bar{\imath}\bar{\jmath}}\,\tilde{\psi}^{\bar{\imath}} \psi^{\bar{\jmath}} \right)\nonumber\\
&&{}
+\nabla^mV_m(\phi,\bar{\phi},\psi,\bar{\psi},F,\bar{F})
\;,
\label{susyL}
\end{eqnarray}
with `holomorphic' superpotential $W(\phi^i)$, $\bar{W}(\bar\phi^{\bar\imath})$, `K\"ahler potential' $K(\phi^i,\bar\phi^{\bar\imath})$,
and the standard notation $K_i\equiv\partial_{\phi^i} K$, $g_{i\bar{\jmath}}\equiv K_{i\bar{\jmath}}$,  $\Gamma^i_{jk}\equiv g^{i \bar\imath} K_{jk\bar\imath}$. 
The background fields $(b^m,M,\bar{M})$ 
have no dynamics and no kinetic terms, while the dynamical fields are given by $n$ pairs of chiral multiplets $(\phi^i,\psi^i)$, $(\bar{\phi}^{\bar{\imath}},\psi^{\bar{\imath}})$, $i,\bar{\imath}=1,\dots, n$, together with auxiliary fields $(F^i, \bar{F}^{\bar\imath})$.  
 We have also added a total derivative to the 
Lagrangian, where the vector $V_m$ depends a priori on all dynamical and auxiliary fields.

We emphasize that the complex scalars $\phi^i$, $\bar{\phi}^{\bar{\imath}}$ are independent and not 
related by complex conjugation. Similarly, $\psi^i$, $\psi^{\bar{\imath}}$, are independent pseudoreal Weyl spinors of opposite chirality.
The resulting Lagrangian (\ref{susyL}) is not real, as usual in Euclidean supersymmetry, see e.g.~\cite{Osterwalder:1973dx,Nicolai:1978,vanNieuwenhuizen:1996tv,Cortes:2003zd}.

Under the rigid supersymmetry transformations given by
\begin{eqnarray}\label{st}
\delta\phi^i &=& -\tilde\zeta \psi^i 
\;,\nonumber\\
\delta\bar\phi^{\bar\imath} &=& -\tilde\xi \psi^{\bar\imath}
\;,\nonumber\\
\delta F^i &=& \tilde\xi \gamma^m \nabla_m \psi^i 
-2 \bar{M}\,\tilde{\zeta} \psi^i -b^m \tilde\xi \gamma_m \psi^i
\;,\nonumber\\
\delta \bar{F}^{\bar\imath} &=& -\tilde\zeta \gamma^m \nabla_m \psi^{\bar\imath} 
-2 M\,\tilde{\xi} \psi^{\bar\imath} -b^m \tilde\zeta \gamma_m {\psi}^{\bar\imath}
\;,\nonumber\\
\delta \psi^i &=& \gamma^m \xi\, \partial_m\phi^i + F^i\,\zeta
\;,\nonumber\\
\delta \psi^{\bar\imath} &=& -\gamma^m \zeta \,\partial_m\bar{\phi}^{\bar\imath} + \bar{F}^{\bar\imath}\,\xi
\;,
\end{eqnarray}
the Lagrangian is invariant up to the following total derivative:
\begin{eqnarray}\label{boundary}
\delta{\cal L} &=& \nabla^m\,({\cal I}_m+\delta V_m)\;,
\nonumber\\[1ex]
{\cal I}_m&\equiv&
K_{i\bar\jmath} \left(\tilde\zeta \gamma_{mn} \psi^i \, \partial^n \bar\phi^{\bar\jmath} 
+\tilde\xi \psi^{\bar\jmath}\,\partial_m \phi^{i} 
+\tilde{\zeta}\gamma_m \psi^{\bar\jmath} F^i
\right)
+2  b_m \left( K_{\bar\imath} \,\tilde\xi \psi^{\bar\imath}-K_i \,\tilde\zeta \psi^i\right)
\nonumber\\
&&{}
+2 \left(MK_i -\tfrac14 W_i \right) \tilde\xi\gamma_m\psi^i
- 2 \left( \bar{M} K_{\bar\imath} -\tfrac14 \bar{W}_{\bar\imath} \right) \tilde{\zeta}\gamma_m\psi^{\bar\imath}
\;,
\end{eqnarray}
provided the supersymmetry parameters $\zeta$, $\xi$ satisfy the Killing spinor equations (\ref{kse}). 
Let us further note that the integrability of (\ref{kse}) gives rise to the relations
\begin{eqnarray}\label{int}
\left\{-\tfrac{1}{2}\,R +12\,(b^mb_m+2M\bar{M})-6\,\nabla_m b^m \right\}\zeta
+ 6\,\gamma^m \xi\,\partial_m M &=& 0
\;,
\nonumber\\
\left\{-\tfrac{1}{2}\,R +12\,(b^mb_m+2M\bar{M})+6\,\nabla_m b^m \right\}\xi
- 6\,\gamma^m \zeta\,\partial_m \bar{M} &=& 0
\;,
\end{eqnarray}
which play a crucial role in verifying the invariance 
of the action (\ref{susyL}) under (\ref{st}).

For a given background $(g_{mn}, b^m,M,\bar{M})$, every solution to the system (\ref{kse}) defines a rigid supersymmetry
of the Lagrangian (\ref{susyL}).  
It has been reported in \cite{fs} that the existence of ${\cal N}=4$ independent solutions of (\ref{kse}) results 
in rather strong constraints on the background fields.\footnote{To avoid any confusion, let us restate that in this notation 
by ${\cal N}$ we count the number of supercharges. E.g.\ minimal supersymmetry
in Minkowski space corresponds to ${\cal N}=4$ independent solutions of (\ref{kse}) (usually referred to as ${\cal N}=1$).
In contrast, most of the backgrounds considered in the following only allow for $\mathcal{N}<4$ supercharges.}
More precisely, the background has to satisfy
\begin{eqnarray}\label{2.11}
Mb_m = \bar{M} b_m = 0 = \partial_m M = \partial_m \bar{M}\;,\qquad
\nabla_m b_n = 0
\;,\nonumber\\
R_{mn} + 8 \left(b_mb_n-g_{mn}b^kb_k\right)-12 g_{mn} M \bar{M} = 0\;,\qquad
W_{mnkl}=0
\;.
\end{eqnarray}
In particular, in this case the four-dimensional background 
metric $g_{mn}$ is necessarily conformally flat. The solutions of (\ref{2.11}) have been further studied in \cite{Jia:2011hw}. 
In contrast, the examples we present in this paper also include backgrounds that only allow for $\mathcal{N}<4$ 
independent solutions of (\ref{kse}), in particular geometries with non-vanishing Weyl tensor and backgrounds 
with non-trivial $(b^m,M,\bar{M})$.

Finally, the auxiliary fields $F^i$ and $\bar{F}^{\bar{\jmath}}$ can be integrated out from the Lagrangian (\ref{susyL})
upon using their field equations,
leading to
\begin{eqnarray}
{\cal L} &=&
-g_{i\bar{\jmath}} \,
\partial_m \phi^i \partial^m \bar{\phi}^{\bar{\jmath}} 
+ g_{i\bar{\jmath}} \,\tilde{\psi}^{\bar{\jmath}}\, \gamma^m 
(\nabla_m \psi^i +\Gamma^i_{jk} \psi^j \partial_m \phi^k 
)
\nonumber\\
&&{}
+\left(M K_{ij}-\tfrac14W_{ij} -(MK_k - \tfrac14 W_k)\Gamma^k_{ij}\right) \tilde\psi^i \psi^j 
\nonumber\\
&&{}
+\left(\bar{M} K_{\bar\imath\bar\jmath}-\tfrac14\bar{W}_{\bar\imath\bar\jmath}
- (\bar{M}K_{\bar{k}} -\tfrac14\bar{W}_{\bar{k}})   \Gamma^{\bar{k}}_{\bar\imath\bar\jmath}\right) \tilde\psi^{\bar\imath}\psi^{\bar\jmath}
\nonumber\\
&&{}
+b^m\left(
2K_i \,\partial_m\phi^i -  2K_{\bar\imath}\, \partial_m\bar{\phi}^{\bar\imath} 
- K_{i\bar{\jmath}}\, \tilde\psi^{\bar\jmath} \gamma_m \psi^i
\right)
\nonumber\\
&&{}
-\left(\tfrac16 R
+4 M\bar{M} -4 b_m b^m \right) K
+ 3\left(\bar{M} W +  M \bar{W} \right) 
+4g^{{i\bar{\jmath}}}\left(MK_i-\tfrac14W_i\right)\left(\bar MK_{\bar\imath}-\tfrac14\bar W_{\bar\imath}\right)\nonumber\\
&&{}
+\nabla^mV_m
\;.
\nonumber\\
\label{susyL0}
\end{eqnarray}
The off-shell supergravity Lagrangian is invariant under combined K\"{a}hler-Weyl transformations. Putting the theory on a fixed classical  background as we have done above, generically breaks this invariance.  
As was shown in \cite{fs}, however, the Lagrangian (\ref{susyL}) is invariant 
under the transformations:
\eq{\label{kahl}\spl{
K(\phi,\bar{\phi})&\rightarrow K(\phi,\bar{\phi})+f(\phi)+\bar{f}(\bar{\phi})\\
W(\phi)&\rightarrow W(\phi)+4Mf(\phi)\\
\bar{W}(\bar{\phi})&\rightarrow \bar{W}(\bar{\phi})+4\bar{M}\bar{f}(\bar{\phi})
~,}}
provided the background satisfies:
\eq{\label{kahlback}
-\frac{1}{24}R+b^mb_m+2M\bar{M}=0~,~~\nabla^mb_m=0
~.}
It is known for example that the $\mathcal{N}=4$ AdS$_4$ background indeed satisfies the above  conditions and hence is invariant under transformations (\ref{kahl}). This has far-reaching implications for the target space of the sigma model \cite{Adams:2011vw}: a simple argument shows that in this case 
the K\"{a}hler form of the target space is exact (assuming there are no divergences in the scalar potential) which in its turn implies that the target space 
is non-compact.

When the background is a Riemannian manifold, 
which is the case we are considering here, the fields $\phi^i$ and $\bar{\phi}^{\bar\imath}$ are not 
related by complex conjugation; the transformations (\ref{kahl}) are not 
strictly-speaking K\"{a}hler transformations and 
the previous argument concerning the exactness of the 
K\"{a}hler form does not go through.

Conditions (\ref{kahlback}) are closely related to the integrability
conditions (\ref{int}). Indeed, by using the methods of section \ref{s22} below,
(\ref{kahlback}) can be seen to be equivalent to the following set
of equations:
\eq{\spl{\label{cloclo}
u\cdot\partial M=u\cdot\partial\bar{M}&=0\\
-\frac{1}{24}R+b_m b^m+2M \bar{M}&=\frac14 e^{A-B}v\cdot\partial\bar{M}
-\frac14 e^{B-A}v^*\cdot\partial{M}\\
\nabla^m b_m&=\frac12 e^{A-B}v\cdot\partial\bar{M}
+\frac12 e^{B-A}v^*\cdot\partial{M}
~,}}
where $A$, $B$ and  $u$, $v$ are defined below in (\ref{etachidef}) and (\ref{uvdef}) respectively.
Hence if the background obeys conditions (\ref{kahlback}) for invariance under
euclidean `K\"{a}hler' transformations (\ref{kahl}), it follows from (\ref{cloclo}) that
\eq{\label{what}
u\cdot\partial M=v^*\cdot\partial{M}=0~,~~u\cdot\partial\bar{M}=v\cdot\partial\bar{M}=0
~.}
Conditions (\ref{what}) can be thought of as locally imposing the `holomorphy' of $M$, $\bar{M}$  with respect to two suitable almost complex structures on the four-dimensional Riemannian manifold $\mathcal{M}_4$.\footnote{$\mathcal{M}_4$ need not admit a global almost
complex structure, as for example in the case $\mathcal{M}_4=S^4$ reviewed
in section \ref{ts}.}

\section{Trivial $G$-structure}\label{s22}

In the following, we will analyze the set of Killing-spinor equations (\ref{kse}) and its solutions, using a suitable $G$-structure;
this section closely follows section B.1 of \cite{lpt}.

For generic backgrounds, the pair of Weyl spinors $\zeta$, $\xi$ which enter 
the Killing-spinor equations are both locally non-vanishing; we may parameterize them as follows:
\eq{\label{etachidef}
\zeta=e^A~\!\eta~,~~~ \xi=e^B~\! \chi
~,
}
where $\eta$, $\chi$ are unimodular Weyl spinors of 
opposite chirality. Moreover we can choose without loss of generality the phases of  $\eta$, $\chi$ so that $A$, $B\in\mathbb{R}$.

For the purposes of the following analysis it will be convenient to 
assume that  $\eta$, $\chi$  are commuting; this 
we are free to do since  the Killing-spinor equations (\ref{kse}) are linear.
Note however that the fermions appearing in the Lagrangian (\ref{l}) are anticommuting.

\subsection*{From spinors to forms}

The pair of unimodular Weyl spinors  $\eta$, $\chi$ locally trivializes the tangent bundle of $\cm_4$, so that on open sets the structure group reduces to $\bbone$. This can also be seen by constructing a pair of complex vectors:
\begin{equation}\label{uvdef}
u^m=\widetilde{\eta}\gamma^m\chi~;~~~
v^m=\widetilde{\eta}\gamma^m\chi^c
~.
\end{equation}
As can be proven by Fierzing, the four real vectors $\mathrm{Re}u$, $\mathrm{Im}u$, $\mathrm{Re}v$, $\mathrm{Im}v$ are unimodular and mutually orthogonal; hence they provide an explicit local trivialization of 
the tangent bundle $T\cm_4$. 

Let us also mention that in deriving the general solution to the Killing spinor equations, it will be  useful to take the following  relations into account:
\begin{equation}\label{useful}\begin{split}
\gamma_m\eta&=v_m\chi-u_m\chi^c\\
\gamma_m\eta^c&=v_{m}^*\chi^c+u_{m}^*\chi\\
\gamma_m\chi&=v_{m}^*\eta+u_m\eta^c\\
\gamma_m\chi^c&=v_m\eta^c-u_{m}^*\eta
~,
\end{split}\end{equation}
which can be shown by Fierzing.

\subsection*{From forms to spinors}\label{formstospinors}\label{s23}

We have seen how one can go from the description in terms of the Weyl 
spinors $\eta$, $\chi$, to a description in terms of the orthonormal frame $u$, $v$ built from the spinor bilinears in (\ref{uvdef}). The converse is also (locally) true: given the orthonormal frame $u$, $v$, one can construct 
the corresponding Weyl spinors  $\eta$, $\chi$, by `inverting' (\ref{useful}). For example, by contracting (\ref{useful}) with the orthonormal frame, we obtain the following projections:
\eq{\spl{
v_m\gamma^m\eta&=u_m\gamma^m\eta=0\\
v_m^{*}\gamma^m\chi&=u_m\gamma^m\chi=0
~.}}
These, together with the unimodularity conditions 
\eq{\eta^\dagger\eta=\chi^\dagger\chi=1~,}
determine $\eta$, $\chi$ up to phase which can then be fixed by 
taking (\ref{uvdef}) into account. This procedure will be carried out  
for the examples in section \ref{examples}, in order to give the 
explicit form of the Killing spinors.

\subsection*{Torsion classes}

The torsion classes of the (trivial) structure of $T\cm_4$ parameterize the failure of $\eta$, $\chi$ to be covariantly constant. Explicitly, we define 
the torsion classes $W_m^{(i)}$, $i=1,\dots 4$, via:
\begin{equation}\begin{split}\label{torsioniia}
\nabla_m\eta&=W_m^{(1)}\eta+W_m^{(2)}\eta^c\\
\nabla_m\chi&=W_m^{(3)}\chi+W_m^{(4)}\chi^c
~,
\end{split}\end{equation}
where $W^{(2,4)}$ are complex one-forms, and $W^{(1,3)}$ are imaginary one-forms; the latter property  follows from the definition (\ref{torsioniia}) upon taking the unimodularity of $\eta$, $\chi$ into account.

Let us also note that alternatively the torsion classes can 
be defined in terms of the exterior derivatives of $u$, $v$. Indeed, from
 eq.~(\ref{torsioniia}) we have, upon taking definition (\ref{uvdef}) into account:
\begin{equation}\begin{split}\label{torsioniiaalt}
\d u&=(W^{(1)}+W^{(3)})\wedge u+W^{(4)}\wedge v-W^{(2)}\wedge v^*\\
\d v&=(W^{(1)}-W^{(3)})\wedge v-W^{(4)*}\wedge u+W^{(2)}\wedge u^*~.
\end{split}\end{equation}

We now proceed by decomposing all forms on the basis of $u$, $v$ -- which can also be thought of as one-forms given the existence of a metric on $\cm_4$.\footnote{In the following we will use the same notation for  both the vectors and the one-forms.} Explicitly, for $i=1,\dots,4$ we decompose:
\begin{equation}\begin{split}
W^{(i)}=\tfrac{1}{2}(u^*W^{(i)}_u
+v^*W^{(i)}_v+uW^{(i)}_{u^*}+vW^{(i)}_{v^*})
~,
\end{split}\end{equation}
where $W^{(i)}_u$ $W^{(i)}_v$, $W^{(i)}_{u^*}$,  $W^{(i)}_{v^*}$ are complex scalars such that $W^{(i)}_u=u\cdot W^{(i)}$, etc. Moreover, the fact that $W^{(1,3)}$ are imaginary implies:
\eq{
W^{(i)}_{u^*}=-W^{(i)*}_u~;~~~
W^{(i)}_{v^*}=-W^{(i)*}_v
~,}
for $i=1,3$.

Taking the above decompositions into account, eqs.~(\ref{torsioniiaalt}) 
can be rewritten as:
\begin{equation}\begin{split}\label{torsioniiaaltalt}
2\d u=&-(W^{(1)}_u+W^{(3)}_u)u\wedge u^*-(W^{(1)}_v+W^{(3)}_v+W^{(2)}_{u^*})u\wedge v^*\\
&+(W^{(1)}_v+W^{(3)}_v+W^{(4)*}_{u^*})^*u\wedge v
-W^{(4)}_u v\wedge u^*
-(W^{(4)}_v+W^{(2)}_{v^*}) v\wedge v^*
-W^{(2)}_uu^*\wedge v^*\\
2\d v=&-(W^{(1)}_u-W^{(3)}_u-W^{(2)}_{v^*})v\wedge u^*
-(W^{(1)}_u-W^{(3)}_u-W^{(4)}_{v})^*u\wedge v\\
&-(W^{(1)}_v-W^{(3)}_v)v\wedge v^*
+(W^{(2)}_{u^*}+W^{(4)*}_{u^*}) u\wedge u^*
-W^{(2)}_vu^*\wedge v^*+W^{(4)^*}_{v^*}u\wedge v^*
~.
\end{split}\end{equation}

\subsection*{Recasting the Killing spinor equations}

Similarly to the decompositions 
for the torsion classes, the complex one-form $b$ can be decomposed as:
\begin{equation}\begin{split}
b=\tfrac{1}{2}(u^*b_u
+v^*b_v+u b_{u^*}+v b_{v^*})
~,
\end{split}\end{equation}
where $b_u$, $b_v$, $b_{u^*}$, $b_{v^*}$, are a priori independent complex scalars. We also need the decompositions of the derivatives of the real scalars $A$, $B$:
\begin{equation}\begin{split}\label{derivsiia}
\d A&=\tfrac{1}{2}\left(u^*(\d A)_u+v^*(\d A)_v+\mathrm{c.c.}\right)\\
\d B&=\tfrac{1}{2}\left(u^*(\d B)_u+v^*(\d B)_v+\mathrm{c.c.}\right)~,
\end{split}\end{equation}
where $(\d A)_u$, $(\d A)_v$, $(\d B)_u$, $(\d B)_v$, are complex scalars.

We are now ready to give the general solution to the Killing spinor equations, by plugging the above expansions into (\ref{kse}), taking eq.~(\ref{useful}) into account.  Explicitly, the Killing spinor equations are equivalent to the following set of conditions:
\boxedeq{\label{217}\spl{
\W^{(1)}_u&=-b_u \\
\W^{(1)}_v&=-2N-b_v\\ 
\W^{(1)}_{{u^*}}&=-3b_{u^*}\\
\W^{(1)}_{{v^*}}&=-3b_{v^*}\\[0.5cm]
W^{(2)}_u&=0\\
W^{(2)}_v&=0\\
W^{(2)}_{{u^*}}&=-2N+2b_v\\
W^{(2)}_{{v^*}}&=-2b_u\\[0.5cm]
\W^{(3)}_u&=b_u\\
\W^{(3)}_v&=3b_v\\
\W^{(3)}_{{u^*}}&=3b_{u^*}\\
\W^{(3)}_{{v^*}}&=2\bar{N}+b_{v^*}\\[0.5cm]
W^{(4)}_u&=0\\
W^{(4)}_v&=-2b_u \\
W^{(4)}_{{u^*}}&=-2\bar{N}+2b_{v^*}\\
W^{(4)}_{{v^*}}&=0~,
}}
where we have defined the complexified torsion classes (recall that 
$W^{(1)}$, $W^{(3)}$ are imaginary):
\begin{equation}\begin{split}
\W^{(1)}:=W^{(1)}+\d A~,~~~
\W^{(3)}:=W^{(3)}+\d B~
\end{split}\end{equation}
and

\begin{equation}\begin{split}
N:=e^{B-A}M~,~~~\bar{N}:=e^{A-B}\bar{M}
~.
\end{split}\end{equation}
We have thus reexpressed the Killing spinor equations, i.e.\ the conditions for 
the background to be supersymmetric, as a set of constraints on the torsion 
classes of the local trivial $G$ structure of $T\mathcal{M}_4$.

The above system of equations can be usefully rewritten in an equivalent way 
as follows:
\begin{equation}\label{altsol}\begin{split}
W^{(2)}_{{u^*}}&=-2N+2b_v\\
\W^{(3)}_u&=b_u\\
\W^{(3)}_v&=3b_v\\
\W^{(3)}_{{u^*}}&=3b_{u^*}\\
\W^{(3)}_{{v^*}}&=2\bar{N}+b_{v^*}\\ 
W^{(4)}_{{u^*}}&=-2\bar{N}+2b_{v^*}\\[0.5cm]
W^{(2)}_u&=0\\
W^{(2)}_v&=0\\
W^{(4)}_u&=0\\
W^{(4)}_{{v^*}}&=0\\
W^{(1)}_{{u}}+W^{(3)}_u&=0\\
W^{(2)}_{{v^*}}-W^{(4)}_{{v}}&=0\\
W^{(2)}_{{u^*}}+W^{(4)*}_{{u^*}}-2(W^{(1)}_{{v}}+W^{(3)}_{v})&=0\\
W^{(4)}_v+2W^{(3)}_u&=-2u\cdot\d B\\
W^{(2)}_{{u^*}}-W^{(4)*}_{{u^*}}&=2v\cdot\d(A+B)\\[0.5cm]
u\cdot\d(A+B)&=0
~.
\end{split}\end{equation}
One strategy for solving the above equations is the following: 
Given a four-manifold $\mathcal{M}_4$ with a specified geometry and an orthonormal frame $u$, $v$ locally trivializing $T^*\mathcal{M}_4$, the 
torsion classes $W^{(i)}$, $i=1,\dots,4$, can be read off of eqs.(\ref{torsioniiaalt}). 
The first six of eqs.~(\ref{altsol}) can then be used to solve for 
$N$, $\bar{N}$ and the four complex components of $b_m$, in terms of the 
torsion classes. The 
remaining ten complex equations then impose constraints on the torsion classes 
and on the derivatives of $A$, $B$.

In other words, given a four-manifold $\mathcal{M}_4$ with a geometry such that 
the last ten of eqs.~(\ref{altsol}) are satisfied, there is no obstruction 
to solving the remaining equations in (\ref{altsol}). Hence the last ten of eqs.~(\ref{altsol}) are necessary and sufficient conditions for obtaining 
a rigid supersymmetric background. Using (\ref{torsioniiaalt}), these necessary and sufficient conditions can be rephrased equivalently in terms of exterior differentials of the orthonormal frame as follows:
\boxedeq{\label{ns}\begin{split}
u\wedge v\wedge \d u&=0\\
u\wedge {v^*}\wedge\d  u&=0\\
v\wedge {v^*}\wedge\d  u&=0\\
u\wedge {v}\wedge\d  v&=0\\
{u^*}\wedge {v}\wedge\d v&=0\\
u\wedge v \wedge\d(e^{2B}{v^*})&=0\\
u \wedge \left[ \mathrm{Im} ({v^*} \wedge\d v) \right]&=0\\
u\wedge v\wedge{v^*}\wedge\d(A+B)&=0\\
v \wedge \left[{v^*}\wedge\d v - \mathrm{Re}({u^*}\wedge\d  u) \right]&=0\\
v \wedge \left[{u^*}\wedge\d u -u\wedge e^{2(A+B)}\d(e^{-2(A+B)}{u^*}) 
\right]&=0
~.
\end{split}}

In section \ref{examples} we will look at several backgrounds which 
satisfy the above conditions.

\section{$SU(2)$ structure}\label{ssu2}

Backgrounds for which $M=0$ or $\bar{M}=0$ allow for one of the two Weyl 
spinors $\zeta$, $\xi$ to vanish identically. In the following we will assume that 
\eq{\bar{M}=0~,}
with a similar analysis for $M=0$. In this case the second of the Killing spinor equations in (\ref{kse}) admits the 
solution
\eq{\xi=0~.\label{z}}
The non-vanishing spinor $\zeta$ can be used to define a local $SU(2)$ structure. Indeed, let us parametrize 
\eq{\zeta=e^A\eta~,}
as in (\ref{etachidef}). 
The unimodular, Weyl spinor $\eta$ defines a local $SU(2)$ structure on $\cm_4$. This can be seen explicitly by constructing a real two-form $J$ and a complex two-form $\omega$ on $\cm_4$ as spinor bilinears:
\eq{\label{jomegadef}
J_{mn}=i\tilde{\eta}\gamma_{mn}\eta^c~;~~~\omega_{mn}=-i\tilde{\eta}\gamma_{mn}\eta
~.
}
The pair $(J,\omega)$ defined above, can be seen by Fierzing to obey the 
definition of an $SU(2)$ structure:
\eq{
J\wedge\omega=0~;~~~J\wedge J=\frac12 \omega\wedge\omega^*\neq0
~.}
On $\cm_4$ there is an almost complex structure, which can be given explicitly in terms of the projectors:
\eq{
\left(\Pi^\pm\right)_m{}^n:=\frac12\left( \delta_m{}^n\mp i J_m{}^n\right)
~.}
Any one-form $V$ can thus be decomposed into (1,0) and  (0,1) parts $V^+$, $V^-$ with respect to the almost complex structure via: 
\eq{\label{pi}V_m^\pm:=\left(\Pi^\pm\right)_m{}^n V^{\ }_n~.}

For our definition of torsion classes we follow closely appendix B of \cite{lpt}. We define 
the torsion classes ${W}_m^{(i)}$, $i=1,2$, via:
\begin{equation}\begin{split}\label{torsioniib}
\nabla_m\eta&={W}_m^{(1)}\eta+{W}_m^{(2)}\eta^c
~,
\end{split}\end{equation}
where as before ${W}^{(2)}$ is a complex one-form, and ${W}^{(1)}$ is an imaginary one-form. Alternatively the torsion classes can be defined in terms of the exterior derivatives of $J$, $\omega$. 
Indeed, from eq.~(\ref{torsioniib}) we have, upon taking definition (\ref{jomegadef}) into account:
\eq{\spl{\label{torsioniibalt}
\d J &= {W}^{(2)*}\wedge\omega+{W}^{(2)}\wedge\omega^* \\
\d \omega &=2{W}^{(1)}\wedge\omega-2{W}^{(2)}\wedge J~.
}}

As already mentioned, the spinor $\eta$ further reduces the structure of $T\cm_4$ from  $Spin(4)\cong SU(2)\times SU(2)'$ (which is accomplished by the 
existence of a Riemannian metric on $\cm_4$) to $SU(2)$. The spinors $\eta$, $\eta^c$ are singlets under the first $SU(2)$ factor, whereas they transform as an $SU(2)'$ doublet under the second factor. Moreover there is an alternative $SU(2)'$-covariant description of the $SU(2)$ structure on $T\cm_4$ and its associated torsion classes, which can be seen as follows: Let us define a  triplet of real two-forms $J_i$, and a triplet of 
real one-forms $\mathcal{W}_i$, $i=1,2,3$, via 
\eq{
(J_1,J_2,J_3):=(J,\mathrm{Re}\omega,
-\mathrm{Im}\omega)
~;~~~
(\mathcal{W}_1,\mathcal{W}_2,\mathcal{W}_3):=(\mathrm{Im}{W}^{(1)},\mathrm{Im}{W}^{(2)},
-\mathrm{Re}{W}^{(2)})
~.}
It can be seen that the $J_i$'s transform as a triplet of $SU(2)'$, and moreover eqs.~(\ref{torsioniibalt}) can be cast in an $SU(2)'$-covariant form:
\eq{
\d J_m=2\varepsilon_{mnp}\mathcal{W}_n \wedge J_p
~.}
We may use this $SU(2)'$ gauge freedom to rotate the torsion classes in eq.~(\ref{torsioniibalt}) to a more standard form, as in \cite{gauntlett}.

In terms of the $SU(2)$ structure the remaining Killing spinor equation, the first line of (\ref{kse}), can be reformulated equivalently as the following 
set of constraints on the torsion classes:
\boxedeq{\spl{\label{su2torsion}
0&=W^{(1)+}+dA^+ +3b^+\\
0&=W^{(1)-}+dA^- +b^-\\
0&=W^{(2)+}-i\omega \cdot b\\
0&=W^{(2)-}
~.}}
In the above $dA^\pm$, $W^{(i)\pm}$, $b^\pm$, are all defined as 
in (\ref{pi}); $\omega \cdot b$ is a shorthand for 
$dx^m\omega_{mn} b^n$.

Equations (\ref{su2torsion}) can be compared to the ones derived in section \ref{s22} as follows. We introduce an auxiliary unimodular Weyl spinor $\chi$ of opposite 
chirality to $\eta$. The pair $(\eta,\chi)$ defines a local trivialization 
which permits us to recast the $SU(2)$ structure in terms of the local orthonormal coframe $(u,v)$ introduced previously. Explicitly, the equations (\ref{jomegadef}) become:
\eq{
J=\frac{i}{2}(u\wedge u^*+v\wedge v^*)~;~~~\omega=-iu\wedge v
~.}
Moreover, it can be seen that the equations in (\ref{su2torsion}) are identical to the first half of the equations in (\ref{217}) for the case where $N=0$.

\subsection*{Global considerations}

As we have already emphasized the construction of the $G$-structure is local, both in the trivial and in the $SU(2)$ case. In particular the existence of local Killing spinors does not imply their global existence. A well-known example is the case of Killing spinors in hyperbolic spaces: although 
Killing spinors can be constructed on $H^n$ \cite{hyper}, they do not survive 
globally on compact quotients $H^n/\Gamma$, where $\Gamma$ is a discrete 
subgroup of $SO(1,n)$.

After constructing the supersymmetric solution using the local $G$-structure approach, one would have to check whether or not the solution can be extended 
globally. This means in particular that one would have to check that 
the Dirac spinor $\zeta+\xi$ can be extended to a global section of the Dirac bundle over $\mathcal{M}_4$.

\section{Examples}\label{examples}

Let us now illustrate the method using explicit examples of four-manifolds. 
The first example we consider is that of a K3 surface. 
Moreover, we will consider backgrounds of the form $\mathcal{M}_4=T^d\times S^{4-d}$ and 
$\mathcal{M}_4=T^d\times H^{4-d}$, for $d=0,1,2$ (the cases $d=3,4$ will not be considered since they lead to flat four-dimensional space). 
In all the cases except for $T^2\times S^2$ the solution extends globally on $\mathcal{M}_4$, i.e.\ all these backgrounds possess global Killing spinors. In the case of $T^2\times S^2$ the background fields develop singularities at 
the poles. Treating the poles as a boundary, taking into consideration the total derivatives,  the `bulk' action can be shown to be supersymmetric. 

Since the topology of hyperbolic space $H^d$ is that of a $d$-dimensional ball, its boundary is a $(d-1)$-dimensional sphere. It follows that for the examples of the form $\mathcal{M}_4=T^d\times H^{4-d}$ the supersymmetry variation of the Lagrangian contains boundary contributions (total derivatives) in general. In the following we will simply assume that the dynamical fields in the Lagrangian (\ref{susyL}) vanish sufficiently fast at the boundary so that the action remains supersymmetric.

In section \ref{sec:gen} we will consider the background $\mathcal{M}_4=T^2\times \mathcal{M}_2$ for general two-dimensional Riemannian manifolds $\mathcal{M}_2$. 
We will show that the necessary and sufficient conditions are satisfied, implying the {local} existence of solutions to the Killing spinor equations. Section \ref{sec:genff} considers a conformally flat $\mathcal{M}_4$.

The backgrounds $T^2\times S^2$ and $T^2\times H^2$ have a non-vanishing Weyl tensor and do not belong to the list of examples considered explicitly in \cite{fs}. Moreover for the $S^1\times S^3$, the $S^1\times H^3$, as well as the example in section \ref{sec:genff}, all of which have a vanishing Weyl tensor, we will present solutions to the Killing spinor equations 
which violate the conditions (\ref{2.11}) and therefore only allow for $\mathcal{N}<4$ supersymmetries.

\subsection*{Scale transformations} 

In all the examples that follow, we fix the overall `radius' $L$ of the 
four-dimensional metric to $L=1$. In order to reinstate the scale $L$ it suffices to perform the following redefinitions:

$\bullet$ {\it four-dimensional metric}: $\d s^2\longrightarrow L^2 \d s^2$

$\bullet$ {\it vierbein}: $e_m{}^a\longrightarrow L e_m{}^a$

$\bullet$ {\it orthonormal frame}: $(u,v)\longrightarrow L(u,v)$

$\bullet$ {\it background fields}: $(M,\bar{M})\longrightarrow L^{-1}(M,\bar{M})$

Everything else: the coordinates $x^m$, the Killing spinors $\zeta$, $\xi$, the warp factors $A$, $B$, the background field $b$ and the torsion 
classes $W^i$, $i=1,\dots,4$, all stay invariant. Of course the individual components of the invariant 
one-forms: $b_u, W_u^i$ etc,  scale like $L^{-1}$.

\subsection{$\mathcal{M}_4=K3$}\label{k3}

This is the most straightforward solution to the Killing spinor equations (\ref{kse}) and is a special case of the class of solutions with $SU(2)$ structure of section \ref{ssu2}. 
It is obtained by setting one of the two spinors to zero, $\xi=0$ as in (\ref{z}), 
while taking the spinor of the opposite chirality to be the 
covariantly constant spinor of the K3 surface:
\eq{
\nabla_m\zeta=0~.
}
The warp factor $A$ is constant, the background fields $b$, $\bar{M}$ and all torsion classes vanish identically so that the equations (\ref{su2torsion}) are 
trivially satisfied. The integrability conditions (\ref{int}) are also identically satisfied, as of course they should, by virtue of the Ricci-flatness of K3. Finally, we note that $M$ remains an a priori unconstrained background scalar in the Lagrangian (\ref{susyL}).

Let us also mention that 
this example trivially satisfies the conditions (\ref{kahlback}) and is therefore invariant under the euclidean `K\"{a}hler' transformations (\ref{kahl}).

\subsection{$\mathcal{M}_4=T^d\times S^{4-d}$}\label{ts}

In this section we consider backgrounds of the form $\mathcal{M}_4=T^d\times S^{4-d}$, for $d=0,1,2$. The case $d=0$ is well-known and 
belongs to the examples presented in \cite{fs}; we mention it here for completeness and in order to facilitate comparison 
with different conventions in the literature. For the cases $d=1,2$ we will present solutions to the Killing spinor equations which violate the conditions (\ref{2.11}) and therefore only allow for $\mathcal{N}<4$ supersymmetries. 
Let us also mention that, as it is easy to check, the cases $d=0,1$ (but not the case $d=2$) satisfy the conditions (\ref{kahlback}) and are therefore invariant under the euclidean `K\"{a}hler' transformations (\ref{kahl}).

\subsection*{$\mathcal{M}_4=S^4$}

In this case the line element of $\mathcal{M}_4$ reads:
\eq{\label{cfmr}
ds^2=\d\theta_4^2+\sin\theta_4 \d\theta_3^2
+\sin\theta_4\sin\theta_3 \d\theta_2^2
+\sin\theta_4\sin\theta_3\sin\theta_2 \d\theta_1^2~,
}
with the orthonormal frame given by
\eq{\spl{\label{21r}
u&=e^{-i\theta_1}\sin\theta_4\big[
(\cos\theta_3+i\sin\theta_3\cos\theta_2)\sin\theta_3\sin\theta_2 \d\theta_1\\
&~~~~~~~~~~~~~~~~~~~~
~~~~~~~~-(\sin\theta_3-i\cos\theta_3\cos\theta_2)\sin\theta_3\d\theta_2
+i\sin\theta_2\d\theta_3
\big]\\
v&=-i\d\theta_4+\sin\theta_4\left[
\cos\theta_2 \d\theta_3
-\sin\theta_3\sin\theta_2 (\cos\theta_3\d\theta_2
+\sin\theta_3\sin\theta_2\d\theta_1)
\right]
~.}}
It is then straightforward to see that the necessary and sufficient conditions 
(\ref{ns}) are indeed satisfied if we take:
\eq{\label{rr} e^A=\sqrt{2}\cos\frac{\theta_4}{2}~;
~~~e^B=\sqrt{2}\sin\frac{\theta_4}{2}~.}
The torsion classes of $\mathcal{M}_4$ can be read off using (\ref{21r}), (\ref{torsioniiaaltalt}):
\eq{
\spl{
W^1&=  i \sin^2\frac{\theta_4}{2} 
   \big[\cos\theta_2\d\theta_3 -\sin\theta_2 \sin\theta_3 (\sin\theta_2 \sin\theta_3\d\theta_1+\cos\theta_3\d\theta_2)\big]\\
W^2&=-e^{-i\theta_1} \sin^2\frac{\theta_4}{2}
   \Big[\sin\theta_2\d\theta_3 \\
&~~~+\sin\theta_3
   \big([\cos\theta_2 \sin
   \theta_3-i \cos\theta_3]\d\theta_1 \sin\theta_2 +[\cos\theta_2\cos\theta_3+i \sin\theta_3]\d\theta_2 \big) \Big]\\
W^3&= i \cos^2\frac{\theta_4}{2} 
   \big[\cos\theta_2\d\theta_3 -\sin\theta_2 \sin\theta_3 (\sin\theta_2 \sin\theta_3\d\theta_1+\cos\theta_3\d\theta_2)\big] \\
W^4&= e^{-i\theta_1} \cos^2\frac{\theta_4}{2}
   \Big[\sin\theta_2\d\theta_3 \\
&~~~+\sin\theta_3
   \big([\cos\theta_2 \sin
   \theta_3-i \cos\theta_3]\d\theta_1 \sin\theta_2 +[\cos\theta_2\cos\theta_3+i \sin\theta_3]\d\theta_2 \big) \Big]~.
}
}
Moreover from the first six equations in (\ref{altsol}) we can determine the background fields:
\eq{\label{bcksolr}
M=-\bar{M}=-\frac{i}{2}~,~~b=0~.
}

As described in section \ref{formstospinors}, from the above we can also 
read off the explicit form of the Killing spinors obeying (\ref{kse}). 
We will use the explicit gamma matrix basis (\ref{explit1}, \ref{explit2}),  
while the coordinate system is given by $(x^1,\dots,x^4)=(\theta_1,\dots,\theta_4)$. With 
these conventions, the Killing spinors are given by:
\eq{\label{k}
\zeta=\left( \begin{array}{c}
e^{-\frac{i}{2}(\theta_1-\theta_3)} \cos
  \frac{\theta_2}{2}  \cos \frac{\theta_4}{2}\\
-ie^{-\frac{i}{2}(\theta_1+\theta_3)} 
   \sin\frac{\theta_2}{2} \cos \frac{\theta_4}{2}\\
0\\
0\end{array} \right)~,~~~
\xi=\left( \begin{array}{c}
0\\
0\\
e^{-\frac{i}{2}(\theta_1-\theta_3)} \cos
  \frac{\theta_2}{2}  \sin \frac{\theta_4}{2}\\
-ie^{-\frac{i}{2}(\theta_1+\theta_3)} 
   \sin\frac{\theta_2}{2} \sin \frac{\theta_4}{2}
\end{array} \right)
~.}
This result can also be seen directly from the Killing spinor 
equations (\ref{kse}), by taking (\ref{bcksolr}) into account. The above expressions are identical to the ones for the Killing spinors constructed explicitly  in \cite{lpr}\footnote{This example possesses extended $\mathcal{N}=4$ supersymmetry so there is in fact a four-dimensional space of Killing spinors. The expressions in (\ref{k}) correspond to the particular choice $\epsilon_0=(1,0,0,0)$ in \cite{lpr} ({\it cf.} appendix B therein).}.

\subsection*{ $\mathcal{M}_4=S^1\times S^3$ }

In this case the line element of $\mathcal{M}_4$ reads:\footnote{It is well-known (see e.g. \cite{fs}) that this geometry admits $\mathcal{N}=4$ supersymmetries. This is achieved by taking:
\eq{
M=\bar{M}=0~,~~ b=\frac12\d x~.
}
The Killing spinors obeying (\ref{kse}) then read:
\eq{
\zeta=e^{-x}\left( \begin{array}{c}
c_1e^{-\frac{i}{2}(\theta_1+\theta_3)} \cos
  \frac{\theta_2}{2}-i c_2 e^{\frac{i}{2}(\theta_1-\theta_3)} \sin
  \frac{\theta_2}{2}\\
c_2e^{\frac{i}{2}(\theta_1+\theta_3)} \cos
  \frac{\theta_2}{2}-i c_1 e^{-\frac{i}{2}(\theta_1-\theta_3)} \sin
  \frac{\theta_2}{2}\\
0\\
0\end{array} \right)~,~~~
\xi=e^x\left( \begin{array}{c}
0\\
0\\
c_3e^{-\frac{i}{2}(\theta_1+\theta_3)} \cos
  \frac{\theta_2}{2}-i c_4 e^{\frac{i}{2}(\theta_1-\theta_3)} \sin
  \frac{\theta_2}{2} \\
c_4e^{\frac{i}{2}(\theta_1+\theta_3)} \cos
  \frac{\theta_2}{2}-i c_3 e^{-\frac{i}{2}(\theta_1-\theta_3)} \sin
  \frac{\theta_2}{2}
\end{array} \right)
}
where $c_1,\dots, c_4$ are arbitrary constants. 
Note that these are not periodic in $x$ and hence not globally-defined. 
As is explained in \cite{fs}, this problem can be circumvented by 
using the formalism of `new minimal supergravity' \cite{nms}, and 
therefore solving a modified version of the Killing spinor equations. Here we will present instead a different background with $\mathcal{N}<4$ supersymmetry.}
\eq{\label{cfmtaa}
ds^2=d\theta_3^2+\sin^2\theta_3\d\theta_2^2
+\sin^2\theta_3\sin^2\theta_2\d\theta_1^2+ \d x^2~.
}
The orthonormal frame is given by
\eq{\spl{\label{21taa}
u&=-e^{-i\theta_1}\big[
(\cos\theta_3+i\sin\theta_3\cos\theta_2)\sin\theta_3\sin\theta_2 \d\theta_1\\
&~~~~~~~~~~~~~~~~~~~~
~~~~~~~~-(\sin\theta_3-i\cos\theta_3\cos\theta_2)\sin\theta_3\d\theta_2
+i\sin\theta_2\d\theta_3
\big]\\
v&=i\d x-\cos\theta_2 \d\theta_3
+\sin\theta_3\sin\theta_2 (\cos\theta_3\d\theta_2
+\sin\theta_3\sin\theta_2\d\theta_1)
~.}}
It is then easy to see that the necessary and sufficient conditions 
(\ref{ns}) are indeed satisfied if we take:
\eq{\label{rtaa} A=B=0~.}
The torsion classes of $\mathcal{M}_4$ can be read off using (\ref{21taa}), (\ref{torsioniiaaltalt}):
\eq{
\spl{
W^1&= W^3=-\frac{i}{2} \mathrm{Re}(v)\\
W^2&=- W^4=-\frac{i}{2}u~.
}
}
Moreover from the first six equations in (\ref{altsol}) we can determine the background fields:
\eq{\label{bcksoltaa}
M=-\bar{M}=\frac{i}{3}~,~~b=-\frac{1}{6}\d x~.
}

As described in section \ref{formstospinors}, from the above we can also 
read off the explicit form of the Killing spinors obeying (\ref{kse}). 
We will use the explicit gamma matrix basis (\ref{explit1}, \ref{explit2}),  
while the coordinate system is given by $(x^1,x^2,x^3,x^4)=(\theta_1,\theta_2,\theta_3,x)$. With 
these conventions, the Killing spinors are given by:
\eq{
\zeta=\left( \begin{array}{c}
e^{-\frac{i}{2}(\theta_1-\theta_3)} \cos
  \frac{\theta_2}{2} \\
-ie^{-\frac{i}{2}(\theta_1+\theta_3)} 
   \sin\frac{\theta_2}{2} \\
0\\
0\end{array} \right)~,~~~
\xi=\left( \begin{array}{c}
0\\
0\\
-e^{-\frac{i}{2}(\theta_1-\theta_3)} \cos
  \frac{\theta_2}{2} \\
ie^{-\frac{i}{2}(\theta_1+\theta_3)} 
   \sin\frac{\theta_2}{2} 
\end{array} \right)
~.}
This result can also be seen directly from the Killing spinor 
equations (\ref{kse}), by taking (\ref{bcksoltaa}) into account.
Let us finally note that although this background is conformally flat,
the background fields (\ref{bcksoltaa}) do not satisfy the conditions (\ref{2.11}),
showing that this background does not admit ${\cal N}=4$ unbroken supersymmetries, although an ${\cal N}=2$ supersymmetry can be made manifest.

\subsection*{ $\mathcal{M}_4=T^2\times S^2$ }

In this case the line element of $\mathcal{M}_4$ reads:
\eq{\label{cfmtt}
ds^2=d\theta^2+\sin^2\theta\d\varphi^2+ \d x^2+\d y^2~,
}
with the orthonormal frame given by
\eq{\label{21tt}
u=\d x+i\d y~, ~~
v=\d\theta +i\sin\theta\d\varphi
~.}
It is then easy to see that the necessary and sufficient conditions 
(\ref{ns}) are indeed satisfied if we take:\footnote{More generally we could consider solutions of the equation $A+B=0$, which leaves 
one of the functions $A$, $B$ undetermined; the general case will 
be treated in section \ref{sec:gen}.}
\eq{\label{rtt} A=B=0~.}
The torsion classes of $\mathcal{M}_4$ can be read off using (\ref{21tt}), (\ref{torsioniiaaltalt}). The only non-zero ones are:
\eq{
\spl{
W^1&=-W^3=-\frac{i}{2}\cos\theta\d\varphi~.
}
}
Moreover from the first six equations in (\ref{altsol}) we can determine the background fields:
\eq{\label{bcksoltt}
M=-\bar{M}=-\frac16\cot\theta~,~~b=\frac{i}{6}\cos\theta\d\varphi~.
}

As described in section \ref{formstospinors}, from the above we can also 
read off the explicit form of the Killing spinors obeying (\ref{kse}). 
We will use the explicit gamma matrix basis (\ref{explit1}, \ref{explit2}),  
while the coordinate system is given by $(x^1,x^2,x^3,x^4)=(x,y,\varphi,\theta)$. With 
these conventions, the Killing spinors are given by:
\eq{\label{kj}
\zeta=e^{\frac{i\pi}{4}}\left( \begin{array}{c}
1\\
0\\
0\\
0\end{array} \right)~,~~~
\xi= e^{-\frac{i\pi}{4}}\left( \begin{array}{c}
0\\
0\\
1\\
0\end{array} \right)
~.}
This result can also be seen directly from the Killing spinor 
equations (\ref{kse}), by taking (\ref{bcksoltt}) into account.

The description in terms of the 
coordinate system $(\varphi$, $\theta)$ breaks down at the north and south poles $\theta=0,\pi$ of $S^2$. In order to verify the supersymmetry of the action, we shall consider the poles as a boundary and examine the variation 
(\ref{boundary}) explicitly, also taking total derivatives into consideration. Substituting (\ref{bcksoltt}) in (\ref{boundary}) we obtain:
\eq{\spl{\label{ztr}
\delta S&=\int_0^{2\pi}\d\varphi \int_0^{\pi}\d\theta ~  \Big{\{} \frac{i}{3} \cot\theta\partial_{\varphi}\left[
K_{\bar\imath}\tilde{\xi}\psi^{\bar\imath}-K_{i}\tilde{\zeta}\psi^{i}
+i(K_{i}\tilde{\xi}\gamma_3\psi^{i}+K_{\bar\imath}\tilde{\zeta}\gamma_3\psi^{\bar\imath})
\right] \\
&-\frac{1}{3}\partial_{\theta}\left[\cos\theta(
K_{i}\tilde{\xi}\gamma_4\psi^{i}+K_{\bar\imath}\tilde{\zeta}\gamma_4\psi^{\bar\imath}
)\right]+  \partial_{\varphi}\Big[\frac{1}{\sin\theta}\delta V_{\varphi}
\Big] +\partial_{\theta}\Big[\sin\theta
\delta V_{\theta}
\Big] 
+\cdots~ \Big{\}},}}
where the ellipses stand for terms which do not depend on the background fields $(M, \bar{M}, b^m)$. Moreover, taking into account the supersymmetry variations (\ref{st}) and the identities
\eq{
\tilde{\zeta}=\tilde{\xi}\gamma_4=i\tilde{\xi}\gamma_3~,~~~
\tilde{\xi}=\tilde{\zeta}\gamma_4=-i\tilde{\zeta}\gamma_3
~,}
which follow from (\ref{kj},\ref{explit1},\ref{explit2}), equation (\ref{ztr}) reduces to:
\eq{\spl{\label{ztrrr}
\delta S&=\int_0^{2\pi}\d\varphi \int_0^{\pi}\d\theta ~  \Big{\{}
\partial_{\varphi} \Big[\frac{1}{\sin\theta}\delta V_{\varphi}
\Big] +
\frac{1}{3}
\partial_{\theta} \Big[
\cos\theta \delta K+
3\sin\theta
\delta V_{\theta}\Big]
+\cdots  \Big{\}}~. 
}}
Hence by choosing the vector $V_m$ in the Lagrangian (\ref{susyL}) as follows:
\eq{
V_{\varphi}=0~~,~~~V_{\theta}=-\frac13\cot\theta K
~,}
the `bulk' action (i.e.\ with the north and south poles removed) is supersymmetric, provided the dynamical fields are regular everywhere on the two-sphere.

Finally, let us note that global $U(1)$ rotations of the coframe:
\eq{\label{rot}
u\rightarrow e^{ic}u~~, ~~~v\rightarrow v
~,}
with $c$ a constant phase, leave the metric and the background fields $(M,\bar{M},b^m)$ invariant. 
This introduces a second arbitrary parameter (besides the overall scale of the Killing spinors) to the space of solutions of the 
system of necessary and sufficient conditions (\ref{ns}), making 
an $\mathcal{N}=2$ supersymmetry manifest.
On the other hand, this background is not conformally flat, thus according to (\ref{2.11}) does not allow for 
$\mathcal{N}=4$ supersymmetries.

\subsection{$\mathcal{M}_4=T^d\times H^{4-d}$}\label{th}

In this section we consider backgrounds of the form $\mathcal{M}_4=T^d\times H^{4-d}$, for $d=0,1,2$, where $H^d$ is the hyperbolic space in $d$ dimensions. 
As in the previous section, for the cases $d=1,2$ we will present solutions to the Killing spinor equations which violate the conditions (\ref{2.11})  and therefore only allow for $\mathcal{N}<4$ supersymmetries. Let us also mention that, as it is easy to check, all the examples in this 
section satisfy the conditions (\ref{kahlback}) and are therefore invariant under the euclidean `K\"{a}hler' transformations (\ref{kahl}).

\subsection*{$\mathcal{M}_4=H^4$}

In this case the line element of the hyperbolic space $H^4$ reads:\footnote{This coordinate system covers the entire hyperbolic space. It is related to the Poincaré coordinates 
$$
ds^2=\frac{1}{w^2}\left(\d w^2+\d x^2+\d y^2+\d z^2\right)~,~~~w>0~,
$$
by the coordinate transformation $w=e^{-\rho}$. The boundary of $H^4$ is reached 
at $\rho=\pm\infty$ and has the topology of $S^3$. As already mentioned, we will assume that the dynamical fields in the Lagrangian (\ref{susyL}) vanish sufficiently fast at the boundary so that the action remains supersymmetric.
}
\eq{\label{cfmra}
ds^2=\d\rho^2+e^{2\rho}\left(\d x^2+\d y^2+\d z^2\right)~,~~~x,y,z,\rho\in\mathbb{R}~,
}
with the orthonormal frame given by
\eq{\label{21ra}
u=e^{\rho}(\d x+i\d y)~, ~~
v=\d\rho+i e^{\rho}~\d z
~.}
It is then easy to see that the necessary and sufficient conditions 
(\ref{ns}) are indeed satisfied if we take:
\eq{\label{r} A=B=\frac{\rho}{2}~.}
The torsion classes of $\mathcal{M}_4$ can be read off using (\ref{21ra}), (\ref{torsioniiaaltalt}). The only non-zero ones are:
\eq{
\spl{
W^1&=-W^3=-\frac{i}{2}e^\rho \d z\\
W^2&=-W^{4}=\frac{1}{2}e^\rho (\d x+i\d y)~.
}
}
Moreover from the first six equations in (\ref{altsol}) we can determine the background fields:
\eq{\label{bcksolra}
M=-\bar{M}=-\frac12~,~~b=0~.
}

As described in section \ref{formstospinors}, from the above we can also 
read off the explicit form of the Killing spinors obeying (\ref{kse}). 
We will use the explicit gamma matrix basis (\ref{explit1}, \ref{explit2}),  
while the coordinate system is given by $(x^1,x^2,x^3,x^4)=(x,y,z,\rho)$. With 
these conventions, the Killing spinors are given by:
\eq{
\zeta=e^{\frac{\rho}{2}+\frac{i\pi}{4}}\left( \begin{array}{c}
1\\
0\\
0\\
0\end{array} \right)~,~~~
\xi= e^{\frac{\rho}{2}-\frac{i\pi}{4}}\left( \begin{array}{c}
0\\
0\\
1\\
0\end{array} \right)
~.}
This result can also be seen directly from the Killing spinor 
equations (\ref{kse}), by taking (\ref{bcksolra}) into account. Let us also note that although we have only  manifestly
displayed one supercharge, this example can be shown to possess ${\cal N}=4$ supersymmetry.

\subsection*{ $\mathcal{M}_4=S^1\times H^3$}

As for the case of $S^1 \times S^3$, the model we present here violates the conditions (\ref{2.11}) and hence only admits ${\cal N}<4$ 
supersymmetries although an ${\cal N}=2$ supersymmetry can be made manifest. The line element of $\mathcal{M}_4$ reads:
\eq{\label{cfm3}
ds^2=d\rho^2+e^{2\rho}\left(\d x^2+\d y^2\right)+\d z^2~,
}
with the orthonormal frame given by
\eq{\label{213}
u=e^{\rho}(\d x+i\d y)~, ~~
v=\d\rho+i \d z
~.}
It is then easy to see that the necessary and sufficient conditions 
(\ref{ns}) are indeed satisfied if we take:
\eq{\label{r3} A=B=\frac{\rho}{2}~.}
The torsion classes of $\mathcal{M}_4$ can be read off using (\ref{213}), (\ref{torsioniiaaltalt}). The only non-zero ones are:
\eq{
W^2=-W^{4}=\frac{1}{2}e^\rho (\d x+i\d y)~.
}
Moreover from the first six equations in (\ref{altsol}) we can determine the background fields:
\eq{\label{bcksol3}
M=-\bar{M}=-\frac13~,~~b=-\frac{i}{6}\d z~.
}

As described in section \ref{formstospinors}, from the above we can also 
read off the explicit form of the Killing spinors obeying (\ref{kse}). 
We will use the explicit gamma matrix basis (\ref{explit1}, \ref{explit2}),  
while the coordinate system is given by $(x^1,x^2,x^3,x^4)=(x,y,z,\rho)$. With 
these conventions, the Killing spinors are given by:
\eq{
\zeta=e^{\frac{\rho}{2}+\frac{i\pi}{4}}\left( \begin{array}{c}
1\\
0\\
0\\
0\end{array} \right)~,~~~
\xi= e^{\frac{\rho}{2}-\frac{i\pi}{4}}\left( \begin{array}{c}
0\\
0\\
1\\
0\end{array} \right)
~.}
This result can also be seen directly from the Killing spinor 
equations (\ref{kse}), by taking (\ref{bcksol3}) into account.

\subsection*{ $\mathcal{M}_4=T^2\times H^2$ }

In this case the line element of $\mathcal{M}_4$ reads:
\eq{\label{cfmt}
ds^2=\d r^2+e^{2r}\d z^2+ \d x^2+\d y^2~,
}
with the orthonormal frame given by
\eq{\label{21t}
u=\d x+i \d y~, ~~
v=\d r+i e^{r}\d z
~.}
It is then easy to see that the necessary and sufficient conditions 
(\ref{ns}) are indeed satisfied if we take:
\eq{\label{rt} A=B=0~.}
The torsion classes of $\mathcal{M}_4$ can be read off using (\ref{21t}), (\ref{torsioniiaaltalt}). The only non-zero ones are:
\eq{
\spl{
W^1&=-W^3=-\frac{i}{4} e^{r}\d z~.
}
}
Moreover from the first six equations in (\ref{altsol}) we can determine the background fields:
\eq{\label{bcksolt}
M=-\bar{M}=-\frac16 ~,~~b=\frac{i}{6} e^{r}\d z~.
}

As described in section \ref{formstospinors}, from the above we can also 
read off the explicit form of the Killing spinors obeying (\ref{kse}). 
We will use the explicit gamma matrix basis (\ref{explit1}, \ref{explit2}),  
while the coordinate system is given by $(x^1,x^2,x^3,x^4)=(x,y,z,r)$. With 
these conventions, the Killing spinors are given by:
\eq{
\zeta=e^{\frac{i\pi}{4}}\left( \begin{array}{c}
1\\
0\\
0\\
0\end{array} \right)~,~~~
\xi= e^{-\frac{i\pi}{4}}\left( \begin{array}{c}
0\\
0\\
1\\
0\end{array} \right)
~.}
This result can also be seen directly from the Killing spinor 
equations (\ref{kse}), by taking (\ref{bcksolt}) into account. Finally, let us mention that this example allows for $U(1)$ coframe rotations as in  
eq.~(\ref{rot}) leaving all background fields invariant; for the reasons 
that were previously explained, this shows that the theory is $\mathcal{N}=2$ supersymmetric.

\subsection{$\mathcal{M}_4=T^2\times \mathcal{M}_2$}\label{sec:gen}

Let us now consider the background $\mathcal{M}_4=T^2\times \mathcal{M}_2$ for a general two-dimensional Riemannian manifold $\mathcal{M}_2$. 
In this case the line element of $\mathcal{M}_4$ reads:
\eq{\label{cfmtg}
ds^2=u\otimes u^*+v\otimes v^*~,
}
with $u=\d x_1+i\d x_2$ a complex one-form on $T^2$. Moreover we will 
take $v$ to be independent of the coordinates $x_1$, $x_2$ of $T^2$, so that
\eq{\label{21tg}
\d u=0~;~~~\d v=f v\wedge v^*~,
}
for a function $f$ of $\mathcal{M}_2$.\footnote{Note that $f$ is a scalar 
under diffeomorphisms, but is 
not invariant under $SO(4)$ transformations of the orthonormal frame.} We will further 
assume that  $A$, $B$ are also functions of the coordinates of $\mathcal{M}_2$, i.e.\ independent of  $x_1$, $x_2$. It is then easy to see that the necessary and sufficient conditions 
(\ref{ns}) are indeed satisfied if we take:
\eq{\label{rtg} {A+B}=\mathrm{constant}~.}
Without loss of generality we will henceforth take the right-hand side above to be zero. 
The torsion classes of $\mathcal{M}_4$ can be read off using (\ref{21tg}), (\ref{torsioniiaaltalt}). The only non-zero ones are:
\eq{
\spl{
W^1&=-W^3=i~\! \mathrm{Im}(f^* v)~
}
}
Finally, from the first six equations in (\ref{altsol}) we can determine the background fields:
\eq{\label{bcksoltg}
M=\frac{1}{3}e^{2A}(f-v\cdot\d A)~,
~~\bar{M}=-\frac{1}{3}e^{-2A}(f^*+v^*\cdot\d A)~,~~b=-\frac{1}{3}\left[\d A+i~\! \mathrm{Im}(f^* v)\right]~.
}
Moreover, the form of the solution implies that
\eq{
M \bar{M}-b^mb_m=0~,
}
is satisfied identically.

Explicit expressions for the background fields and the Killing spinors can also be obtained as follows. We can always choose local coordinates so that
\eq{ 
u=\d x_1+i \d x_2~, ~~
v=e^{\phi(x_3,x_4)}(\d x_4+i \d x_3)
~,}
where the function $\phi$ is related to $f$ in (\ref{21tg}) through
\eq{
f=-\frac{1}{2}({\partial_4\phi}
+i {\partial_3\phi})e^{-\phi}
~.}
In this coordinate system the torsion classes read:
\eq{
\spl{
W^1&=-W^3=-\frac{i}{2}(\d x_3{\partial_4\phi}
-\d x_4{\partial_3\phi})~,
}
}
while the background fields are given by:
\eq{\label{bcksoltgg}\spl{
b&=-\frac{1}{6}
\Big[\d x_3(2\partial_3A-{i}\partial_4\phi)
+\d x_4(2\partial_4A+{i}\partial_3\phi)\Big]\\
M&=-\frac{1}{6}e^{2A-\phi}({\partial_4}
+i {\partial_3}) (\phi+2A)~,
~~\bar{M}=\frac{1}{6}e^{-2A-\phi}({\partial_4}
-i {\partial_3}) (\phi-2A)~.
}}
Finally, the Killing spinors read:
\eq{
\zeta=e^{A+\frac{i\pi}{4}}\left( \begin{array}{c}
1\\
0\\
0\\
0\end{array} \right)~,~~~
\xi= e^{-A-\frac{i\pi}{4}}\left( \begin{array}{c}
0\\
0\\
1\\
0\end{array} \right)
~.}
This result can also be seen directly from the Killing spinor 
equations (\ref{kse}), by taking (\ref{bcksoltgg}) into account.

The above analysis guarantees the existence of local Killing spinors on $\mathcal{M}_4$; it is important to note, however, that the global existence is 
not guaranteed a priori.  Finally, let us mention that this example allows for $U(1)$ coframe rotations as in  
eq.~(\ref{rot}) leaving all background fields invariant; for the reasons 
that were previously explained, this shows that the theory is $\mathcal{N}=2$ supersymmetric.

\subsection{$\mathcal{M}_4$ conformally flat}\label{sec:genff}

We finally consider a conformally-flat background $\mathcal{M}_4$ such that the line element reads:
\eq{\label{cfmtgf}
ds^2=e^{2\phi(x_3,x_4)}\left(\d x_1^2+\d x_2^2+\d x_3^2+\d x_4^2\right)~,
}
for  $\phi$ a function of $x_3$, $x_4$. 
The orthonormal frame given by
\eq{\label{21tff}
u=e^{\phi}\left(\d x_1+i \d x_2\right)~, ~~
v=e^{\phi}\left(\d x_4+i \d x_3\right)
~.}
We will further 
assume that  $A$, $B$ are also functions of the coordinates of  $x_3$, $x_4$. It is then easy to see that the necessary and sufficient conditions 
(\ref{ns}) are indeed satisfied if we take:
\eq{\label{rtgff} {A+B}=\phi~.}
The torsion classes of $\mathcal{M}_4$ can be read off using (\ref{21tff}), (\ref{torsioniiaaltalt}). The only non-zero ones are:
\eq{
\spl{
W^1&=-W^3=\frac{i}{2}(\partial_3\phi\d x_4-\partial_4\phi\d x_3)\\
W^2&=\frac12 (\partial_4\phi+i\partial_3\phi)(\d x_1+i\d x_2)\\
W^4&=-\frac12 (\partial_4\phi-i\partial_3\phi)(\d x_1+i\d x_2)~.
}
}
Finally, from the first six equations in (\ref{altsol}) we can determine the background fields:
\eq{\label{bcksoltgff}\spl{
b&=\frac{i}{6}\left[\partial_3(\phi-2A)\d x_4-\partial_4(\phi-2A)\d x_3\right]\\
M&=-\frac{1}{3}e^{2(A-\phi)}(\partial_4+i\partial_3)(A+\phi)~,
~~\bar{M}=\frac{1}{3}e^{-2A}(\partial_4-i\partial_3)(A+\phi)~.
}}
Note that by taking $A=B=\phi/2$ it follows from the equation 
above that we can set $b=0$. However, for general $A$, the one-form $b$ is 
neither zero nor covariantly constant. Indeed, a short calculation gives:
\eq{\label{ncc}
\nabla^mb_m=4(\partial_3\phi\partial_4A-\partial_4\phi\partial_3A)
~.}
We see that for general $A$, eqs.~(\ref{bcksoltgff}), (\ref{ncc}) violate the conditions (\ref{2.11}),
showing that this background does not allow for ${\cal N}=4$ independent supersymmetries.  It does however allow for $U(1)$ coframe rotations as in  
eq.~(\ref{rot}) leaving all background fields invariant;  this shows that the theory is $\mathcal{N}=2$ supersymmetric.

The Killing spinors read:
\eq{
\zeta=e^{A+\frac{i\pi}{4}}\left( \begin{array}{c}
1\\
0\\
0\\
0\end{array} \right)~,~~~
\xi= e^{\phi-A-\frac{i\pi}{4}}\left( \begin{array}{c}
0\\
0\\
1\\
0\end{array} \right)
~.}
As already emphasized in the previous example, the above analysis guarantees the existence of local Killing spinors on $\mathcal{M}_4$ however the global existence is not guaranteed a priori.

\section{Conclusions}

We have presented a systematic approach for the solution of the Killing spinor equations
in four-dimensional Riemannian space, whose solutions define the backgrounds on which
globally supersymmetric field theories can be formulated. 
The general globally supersymmetric Lagrangian has been constructed 
directly in Euclidean signature from the outset, without reference to any Wick rotation
keeping track of potential boundary terms. 
We have reformulated the conditions for rigid supersymmetry in  
Riemannian space in terms of $G$-structures and given explicit expressions for the background fields
in terms of the torsion classes.

We have illustrated the formalism with several explicit examples
which go beyond the list of backgrounds discussed in \cite{fs,Jia:2011hw}.
In particular, our examples 
include four-dimensional backgrounds 
for which the Weyl tensor is non-vanishing, as well as examples with 
vanishing Weyl tensor which violate the conditions (\ref{2.11}) and therefore only allow for $\mathcal{N}<4$ supersymmetries. 
As we have seen, the K3 and all the $T^d\times S^{4-d}$ and $T^d\times H^{4-d}$ examples, except for the case 
of $T^2\times S^{2}$, satisfy the conditions (\ref{kahlback}) and are therefore invariant under the euclidean `K\"{a}hler' transformations (\ref{kahl}).

We expect our analysis and results to be useful in extending the list of known 
rigid supersymmetric theories in curved backgrounds. 
It would also be 
interesting to apply these methods to the study of 
rigid supersymmetric theories in backgrounds of dimension different than four.

\appendix

\section{Spinors and gamma matrices in Euclidean spaces}\label{sec:spinors}

In this section we list some useful relations and 
explain in more detail our spinor conventions for general even-dimensional Euclidean spaces of dimension $2k$. 

The charge conjugation matrix obeys:
\eq{\label{c}
C^{\mathrm{Tr}}=(-)^{\frac{1}{2}k(k+1)}C~;~~C^*=(-)^{\frac{1}{2}k(k+1)}C^{-1}~;~~
\gamma_m^{\mathrm{Tr}}=(-)^k C^{-1}\gamma_m C
~.
}
The complex conjugate $\eta^c$ of a spinor $\eta$ is given by:
\eq{
\eta^c:=C\eta^*
~,}
from which it follows that:
\eq{
(\eta^c)^c=(-)^{\frac12 k(k+1)}\eta
~.}
Covariantly-transforming spinor bilinears must be of the form $(\widetilde{\psi}\gamma_{m_1\dots m_p}\chi)$, 
where in any dimension we define:
\eq{
\widetilde{\psi}:=\psi^{\mathrm{Tr}}C^{-1}
~.}
One can also show the following useful identity:
\eq{
\gamma_{m_1\dots m_p}^*=(-)^{kp}C^{-1}\gamma_{m_1\dots m_p}C
~.}

The case of four-dimensional Euclidean space is obtain by specializing to 
 $k=2$. The chiral irreducible representation of $Spin(4)$ is 
pseudoreal. This means that given a Weyl spinor $\eta$, both 
$\eta$ and its complex conjugate $\eta^c$ have the same chirality.

For the explicit examples of section \ref{examples} we use the 
following flat-space gamma matrix basis:
\eq{\label{explit1}
\gamma_{i}= \left( \begin{array}{cc}
0 & \sigma_i  \\
\sigma_i & 0\end{array} \right)~,~~~
\gamma_{4}= \left( \begin{array}{cc}
0 & i\bbone  \\-i\bbone & 0\end{array} \right)
~,}
where $\sigma_i$, $i=1,2,3$, are the Pauli matrices. Moreover, in 
this basis the chirality and charge-conjugation matrices read:
\eq{\label{explit2}
\gamma_{5}= \left( \begin{array}{cc}
\bbone & 0  \\0 &-\bbone \end{array} \right)
~,~~~
C=\left( \begin{array}{cc}
-i\sigma_2 & 0  \\0 &i\sigma_2 \end{array} \right)
~.}
%

%
%

\end{document}